\documentclass[submission,copyright,creativecommons]{eptcs}
\providecommand{\event}{QSQW 2020} 
\usepackage{breakurl}              
\usepackage{underscore}            
\usepackage{tikz,braket,amsmath,amsfonts,mathdots}
\usetikzlibrary{calc}

\newcommand\id{\mathbb{I}}
\newcommand\hilb{\mathcal{H}}
\newcommand{\Span}{\mathop{\mathrm{Span}}\nolimits}
\newcommand\trho{{\mathord{\sim}_\rho}}
\newcommand{\red}{\text{\$}}

\title{Projection Theorem for Discrete-Time Quantum Walks}
\author{V\'{a}clav Poto\v{c}ek
\institute{Faculty of Nuclear Sciences and Physical Engineering,\\Czech Technical University in Prague,\\B\v{r}ehov\'{a} 7, 115 19, Praha 1, Czech Republic}
\email{vaclav.potocek@fjfi.cvut.cz}}
\def\titlerunning{Projection Theorem for Discrete-Time Quantum Walks}
\def\authorrunning{V. Poto\v{c}ek}
\begin{document}
\maketitle

\begin{abstract}
We make and generalize the observation that summing of probability amplitudes of a discrete-time quantum walk over partitions of the walking graph consistent with the step operator results in a unitary evolution on the reduced graph which is also a quantum walk.
Since the effective walking graph of the projected walk is not necessarily simpler than the original, this may bring new insights into the dynamics of some kinds of quantum walks using known results from thoroughly studied cases like Euclidean lattices.
We use abstract treatment of the walking space and walker displacements in aim for a generality of the presented statements.
Using this approach we also identify some pathological cases in which the projection mapping breaks down.

For walks on lattices, the operation typically results in quantum walks with hyper-dimensional coin spaces.
Such walks can, conversely, be viewed as projections of walks on inaccessible, larger spaces, and their properties can be inferred from the parental walk.
We show that this is is the case for a lazy quantum walk, a walk with large coherent jumps and a walk on a circle with a twisted boundary condition.
We also discuss the relation of this theory to the time-multiplexing optical implementations of quantum walks.
Moreover, this manifestly irreversible operation can, in some cases and with a minor adjustment, be undone, and a quantum walk can be reconstructed from a set of its projections.
\end{abstract}

\section{Introduction}
\label{sec:intro}

The present work started as an attempt at making a very general statement applicable to a broad class of discrete-time quantum walks.
Nevertheless, the field has grown so large that any attempt at an all-encompassing formulation of a ``quantum walk'' model, of which all the known varieties would be special cases, seems to be out of reach.
For this reason, it is necessary to state in the beginning that this work is only concerned with coherent, closed, discrete-time quantum walk with a separate position and coin space.
While some other models like scattering discrete-time quantum walks \cite{Scattering,Szegedy}, taking place on edges of the walking graphs rather than vertices, or staggered quantum walks \cite{Staggered}, not requiring a coin state, can to a large extent be considered equivalent to the coined scenarios, this is not always a trivial correspondence \cite{Higuchi,Staggered}.
In the present contribution, quantum walk, discrete-time quantum walk and the acronym DTQW all refer to a particular formulation of a coined discrete-time quantum walk, which will be formalized in Sec.~\ref{sec:theory}.

The projection theorem presented in this work aims to generalize the following observation.
Consider a DTQW on a 2-dimensional Euclidean lattice with displacements to the right, left, up, and down.
The state space is the tensor product of the position space $\hilb_P = \ell^2(\mathbb{Z})$ with a fixed orthonormal basis $\{\ket{x,y} \mid x,y \in \mathbb{Z}\}$ and a coin space $\hilb_C = \mathbb{C}^4$ with its orthonormal basis denoted $\{\ket{R}, \ket{L}, \ket{U}, \ket{D}\}$ for the four basic directions.
In the total state space $\hilb = \hilb_P \otimes \hilb_C$ we naturally consider the tensor product of the two bases.
The evolution consists of alternated application of the coin operator, which we will for this motivatory example consider homogeneous of the form
\begin{equation}
C = \id_P \otimes C_0, \quad C_0 \in U(4),
\end{equation}
and the step operator
\begin{equation}
S = \sum_{x, y} \big( \ket{x+1,y}\bra{x,y} \otimes \Pi_R + \ket{x-1,y}\bra{x,y} \otimes \Pi_L + \ket{x,y+1}\bra{x,y} \otimes \Pi_U + \ket{x,y-1}\bra{x,y} \otimes \Pi_D \big)
\end{equation}
(where $\Pi_x$ are orthogonal projectors on the rays spanned by the respective vectors $\ket{x}$ in $\hilb_C$), forming together a single step evolution operator $U = SC$.

The evolution operator connects the states between step $t$ and $t+1$,
\begin{equation}
\ket{\psi_{t+1}} = \ket{\psi_t},
\end{equation}
but resolving this equation in the position space basis
\begin{equation}
\ket{\psi_t} = \sum_{x,y\in\mathbb{Z}} \ket{x,y} \otimes \ket{\alpha_{x,y}^{(t)}} \qquad
(\ket{\alpha_{x,y}^{(t)}} \in \mathbb{C}^4, \text{ not normalized}),
  \label{eq:ex:alpha}
\end{equation}
one can write the equivalent recurrence relation
\begin{equation}
\ket{\alpha_{x,y}^{(t+1)}} = \Pi_R C \ket{\alpha_{x-1,y}^{(t)}} + \Pi_L C \ket{\alpha_{x+1,y}^{(t)}} + \Pi_U C \ket{\alpha_{x,y-1}^{(t)}} + \Pi_D C \ket{\alpha_{x,y+1}^{(t)}}
\label{eq:rec2d}
\end{equation}

The idea giving rise to the theory as presented here lies within noticing that \eqref{eq:rec2d} retains its structure when formally summed over one of the coordinates.
Introducing
\begin{equation}
\ket{\beta_x^{(t)}} = \sum_{y\in\mathbb{Z}} \ket{\alpha_{x,y}^{(t)}},
\label{eq:ex:beta}
\end{equation}
the summation of \eqref{eq:rec2d} can be expressed entirely in terms of this new coefficient:
\begin{equation}
\ket{\beta_x^{(t+1)}} = \Pi_R C \ket{\beta_{x-1}^{(t)}} + \Pi_L C \ket{\beta_{x+1}^{(t)}} + \Pi_U C \ket{\beta_{x}^{(t)}} + \Pi_D C \ket{\beta_{x}^{(t)}}.
\label{}
\end{equation}
The same would be the recurrence relation of a different DTQW, taking place on a line, although with a four-dimensional coin space using the same homogeneous transform $C_0$ on the coin register, with a step operator
\begin{equation}
    S' = \sum_{x\in\mathbb{Z}} (\ket{x+1}\bra{x}\otimes\Pi_R +
      \ket{x-1}\bra{x}\otimes\Pi_L +
      \ket{x}\bra{x}\otimes\Pi_U +
      \ket{x}\bra{x}\otimes\Pi_D),
  \label{eq:4d-step-2d}
\end{equation}
i.e., that of a lazy quantum walk \cite{Inui-lazy} but with two lazy coin states ($\ket{U}$ and $\ket{D}$).
Indeed, for any particular values of the position $x$ and time $t$, the coin state $\ket{\beta_x^{(t)}}$, obtained by \eqref{eq:ex:beta} from a state \eqref{eq:ex:alpha} that was a result of $t$ steps of evolution of some initial state $\ket{\psi_0}$, should be identical to that of evolving an initial state
\begin{equation}
\ket{\psi'_0} = \sum_{x\in\mathbb{Z}} \ket{x} \otimes \ket{\beta_x^{(0)}}
\label{}
\end{equation}
by the lazy quantum walk described above, if such state is well defined (non-zero and normalizable).

This observation is not restricted to summations along the $x$ or $y$ coordinates. In the following we provide a formalisation allowing many other generalizations.

\section{Projection theory}
\label{sec:theory}

The example in the previous section gives a brief introduction to the projection theory, which we will strive to generalize here.

Let $X$ denote an abstract set of positions.
We will require that this is at most countably infinite so that the position space
\begin{equation}
\hilb_P = \ell^2(X) = \Span\{ \ket{x} \mid x\in X\}
\label{}
\end{equation}
is separable.
Let $\Gamma$ be an (at most countable) set or multiset of displacements on $X$, which we define as injections from $X$ to $X$.
For a given $c \in \Gamma$, we denote $c(x)$ equivalently as a right action of $c$ on the element $x$, $x\cdot c$.
We span another Hilbert space -- the coin space -- over the elements of $\Gamma$:
\begin{equation}
\hilb_C = \ell^2(\Gamma) = \Span\{ \ket{c} \mid c\in\Gamma \}.
\label{}
\end{equation}
It is understood that both $\{\ket{x}\}$ and $\{\ket{c}\}$ form orthonormal bases of their respective Hilbert spaces.
Lastly, we construct $\hilb = \hilb_P \otimes \hilb_C$ along with its orthonormal basis $\{\ket{x}\otimes\ket{c} \mid x\in X, c\in\Gamma\}$.

On this Hilbert space we prescribe a quantum walk using a coin operator, which we assume of the form
\begin{equation}
  C = \sum_{x \in X} \ket{x}\bra{x} \otimes C_x : \ket{x}\ket{c} \mapsto \ket{x}\otimes(C_x\ket{c})
\label{eq:c-before}
\end{equation}
for some map $C_\circ: X \to U(\hilb_C)$, a step operator
\begin{equation}
  S = \sum_{x \in X} \sum_{c \in \Gamma} \ket{x\cdot c}\bra{x} \otimes \ket{c}\bra{c} : \ket{x}\ket{c} \mapsto \ket{x\cdot c}\ket{c}
\label{eq:s}
\end{equation}
and an initial state $\ket{\psi_0} \in \hilb$, which we leave as a free parameter.

One step of a quantum walk is then given by the unitary operator $U=SC$, that is,
\begin{equation}
\ket{\psi_{t+1}} = SC\ket{\psi_t},
\label{}
\end{equation}
which in $n\in\mathbb{N}$ iterations evolves $\ket{\psi_0}$ into
\begin{equation}
\ket{\psi_n} = (SC)^n \ket{\psi_0}.
\label{}
\end{equation}

We now take a \emph{projection map} to be an arbitrary mapping $\rho: X \to X'$ which is onto and satisfies
\begin{equation}
  \forall x, y \in X, \forall c \in \Gamma, \rho(x) = \rho(y) \Leftrightarrow \rho(x\cdot c) = \rho(y\cdot c).
\label{eq:rho-cond}
\end{equation}
This condition means that $\rho$ is consistent with each $c\in\Gamma$ in the following way: $c$ maps an equivalence class of
\begin{equation}
\trho:\ \forall x, y \in X,\ x \sim_\rho y \Leftrightarrow \rho(x) = \rho(y)
\label{}
\end{equation}
to another equivalence class (or the same), so a mapping $c'$ can be constructed such that $c([x]) \subset [y] \Leftrightarrow c'(\rho(x)) = c'(\rho(y))$.
The $\Leftarrow$ implication of \eqref{eq:rho-cond} then provides that $c'$ is injective on $X'$.

The \emph{projection operator} $\red$ induced on $\hilb$ by $\rho$ is then defined as
\begin{equation}
\red: \hilb \to \hilb' : \ket{x}\ket{c} \mapsto \ket{\rho(x)}\ket{c}
\label{eq:red}
\end{equation}
where
\begin{equation}
\begin{aligned}
\hilb_P' &= \ell^2(X'), \\
\hilb_C' &= \hilb_C, \\
\hilb' &= \hilb_P' \otimes \hilb_C'
\end{aligned}
\label{}
\end{equation}
are constructed in complete analogy with $\hilb_P, \hilb_C, \hilb$.

We say that the quantum walk on $\hilb$ prescribed by $C$, $S$, and $\ket{\psi_0}$ allows projection by $\rho$ iff there exists a coin operator $C'$, a step operator $S'$ and an initial state $\ket{\psi_0'}$ such that
\begin{equation}
  \begin{aligned}
S' \red &= \red S, \\
C' \red &= \red C, \\
\red \ket{\psi_0} &= \ket{\psi_0'}.
  \end{aligned}
\label{eq:commutation}
\end{equation}
Then, trivially,
\begin{equation}
  \ket{\psi'_n} := (S'C')^n \ket{\psi'_0} = (S'C')^n \red\ket{\psi_0} = \red(SC)^n \ket{\psi_0} = \red \ket{\psi_n}.
  \label{eq:evo-mapped}
\end{equation}
and thus the quantum walk evolutions of $\ket{\psi_0}$ under $U=SC$ and that of $\ket{\psi_0'}$ under $U'=S'C'$ have a fixed relation that the latter is an image under $\red$ of the former at any step $n$.

The first condition of \eqref{eq:commutation} is satisfied by the naturally defined step operator
\begin{equation}
  S' = \sum_{x' \in X'} \sum_{c \in \Gamma} \ket{c'(x')}\bra{x'} \otimes \ket{c}\bra{c}.
\label{eq:s'}
\end{equation}
It is straightforward to prove that the second condition can be satisfied by an operator of the form
\begin{equation}
  C' = \sum_{x' \in X'} \ket{x'}\bra{x'} \otimes C'_{x'} : \ket{x'}\ket{c} \mapsto \ket{x'}\otimes(C'_{x'}\ket{c}),
\label{eq:c'}
\end{equation}
with some mapping $C'_\circ: X' \to U(\hilb_C)$ if and only if $C_\circ$ is constant over equivalence classes of $\trho$, and if so, the value of $C'_{[x]}$ is that of $C_x$ for the representative $x$ of $[x]$.
Finally, the mapping of $\ket{\psi_0}$ to $\ket{\psi_0'}$ can potentially present a problem only in two cases.
It can happen that $\rho\ket{\psi_0}$ is the zero vector.
In this case, the whole sequence of states \eqref{eq:evo-mapped} is zeroed and the result is not a quantum walk.
The other pathological case is when $\red$ is undefined on $\ket{\psi_0}$.
This can happen whenever the equivalence classes of $\trho$ are infinite.
Let $(x_n)_{n=1}^\infty$ be a sequence of distinct elements of $X$ that all map to a single $x' = \rho(x_n)$, consider a linear combination
\begin{equation}
\ket{\chi} = \sum_{n=1}^\infty \alpha_n \ket{x_n}.
\label{}
\end{equation}
This is a vector in $\hilb_X$ if $(\alpha_n)_{n=1}^\infty \in \ell^2(\mathbb{N})$.
However, formally applying \eqref{eq:red} on $\ket{\chi}\otimes\ket{\gamma}$ for an arbitrary element $\ket{\gamma}$ of $\hilb_C$ yields
\begin{equation}
\red\ket{\chi}\ket{\gamma} = \left( \sum_{n=1}^\infty \alpha_n \right) \ket{x'}\ket{\gamma}
\label{}
\end{equation}
which is only well defined if $(\alpha_n)_{n=1}^\infty \in \ell^1(\mathbb{N})$.
Since there are sequences which belong to $\ell^2(\mathbb{N})$ but not $\ell^1(\mathbb{N})$, there are vectors on which $\red$ is not defined.

In all other cases, the result $\ket{\psi_0'}$ of $\red\ket{\psi_0}$ is a normalizable ket and its norm is conserved under the unitary operation $U'$, so \eqref{eq:evo-mapped} represents a DTQW on $X'$ with coin $C'$, step $S'$ and the initial state $\ket{\psi_0'}$ (or its normalized version, if required).

The mapping \eqref{eq:red} can further be enriched by a phase factor dependent on $x$ and $c$.While this can be done generally, we will leave the discussion for an extended version of this contribution.
Here we will discuss a special but powerful (as illustrated by examples below) case of the theory where some of the introduced objects have additional specific properties.
In particular, we will require in the following that
\begin{itemize}
  \item $X$ is the underlying set of a group $G$,
  \item $\Gamma$ is a subset of $X$ with $x\cdot c$ a right multiplication in $G$,
  \item $X'$ is the underlying set of $G/H$ where $H$ is a normal subgroup of $G$,
  \item $\rho$ is the natural projection homomorphism from $G$ to $G/H$.
\end{itemize}
We note that any group homomorphism automatically satisfies \eqref{eq:rho-cond}, and in case of the natural homomorphism, the condition that $\rho$ is onto is also trivially satisfied.
We will add one more mapping, $\sigma$, a homomorphism from $G$ to $(\mathbb{R},+)$.
With these requirements, we allow phase-added projection operators of the form
\begin{equation}
\red_\varphi: \hilb \to \hilb': \ket{x}\ket{c} \mapsto e^{i\varphi\sigma(x)} \ket{x}\ket{c}.
\label{eq:phase}
\end{equation}

The commutation relations \eqref{eq:commutation} no longer hold with the above choice of operators $S'$, $C'$ when $\red$ is replaced by $\red_\varphi$:
the operator $S'$ needs to be redefined as
\begin{equation}
S' = \sum_{x' \in X'} \sum_{c \in \Gamma} e^{i\varphi\sigma(c)} \ket{c'(x')}\bra{x'} \otimes \ket{c}\bra{c}.
\label{eq:s'-phase}
\end{equation}

\section{Examples}
\label{sec:examples}

In this section we illustrate several examples of projections and some application of the theory.
The examples are limited to the restrictions posed near the end of the previous section.
We work out the first example in more detail, leaving the analogy in the subsequent ones to the reader.

Here, let us consider a canonical quantum walk on the 2D Euclidean lattice, prescribed by $X = \mathbb{Z}^2$, $\Gamma = \{(+1,0), (-1,0), (0,+1), (0,-1)\}$.
The elements of $\Gamma$ we denote in this order by $\{R, L, U, D\}$.
Let $X' = \mathbb{Z}$ with $\rho$ given by
\begin{equation}
  \rho(x,y) = kx+ly, \quad k,l\in\mathbb{Z}, \quad k\perp l.
  \label{eq:rho-k,l}
\end{equation}
Since $k$ and $l$ are coprime, $u$ and $v$ in $\mathbb{Z}$ can be found such that $uk+vl=1$ by B\'{e}zout's identity and by virtue of the same identity these are necessarily coprime as well.
Any pair $(x,y) \in \mathbb{Z}^2$ can then be uniquely written as $(x,y) = p(u,v) + q(-l,k)$ with $p,q \in \mathbb{Z}$.
The normal subgroup $H = \langle (-l,k) \rangle \triangleleft \mathbb{Z}^2$ has cosets of the form
\begin{equation}
  [(x,y)] = p(u,v) + H,
\label{}
\end{equation}
showing that the factor group $G/H$ is isomorphic to the singly generated additive group $[(u,v)]$, which in turn is isomorphic to $(\mathbb{Z},+)$.
The mapping \eqref{eq:rho-k,l} represents the natural projection homomorphism under these isomorphisms.

\begin{figure}[t]
\centering
\begin{tikzpicture}[scale=0.6,
g1/.style={green!50!black,<->},
g2/.style={brown,<->}]
\path[use as bounding box] (-7.5, -3) rectangle (8.5, 3);
\begin{scope}[shift={(-6,-1.5)}]
\foreach \x in {0,...,3} {
\foreach \y in {0,...,3} {
\draw (\x,\y) circle (0.3);
\ifnum\x<3 \draw[g1] (\x+.3,\y) -- (\x+.7,\y); \fi
\ifnum\y<3 \draw[g2] (\x,\y+.3) -- (\x,\y+.7); \fi
}}
\node at (-.4,1.5) [anchor=east] {$\cdots$};
\node at (3.4,1.5) [anchor=west] {$\cdots$};
\node at (1.5,-.2) [anchor=north] {$\vdots$};
\node at (1.5,3.4) [anchor=south] {$\vdots$};
\end{scope}
\node at (0,0) {$\rightarrow$};
\begin{scope}[shift={(3,0)}]
\foreach \x in {0,...,4} {
\draw (\x,0) circle (0.3);
\ifnum\x<4 \draw[g1] (\x+.3,0) -- (\x+.7,0); \fi
\draw[g2] (\x,0) +(70:.3) .. controls +(70:.5) and +(110:.5) .. +(110:.3);
}
\node at (-.4,0) [anchor=east] {$\cdots$};
\node at (4.4,0) [anchor=west] {$\cdots$};
\end{scope}
\end{tikzpicture}
\caption{Mapping of a 2D Euclidean lattice to the walking graph of a lazy walk on a line. The double arrows represent two directed edges each.}
\label{fig:lazy}
\end{figure}
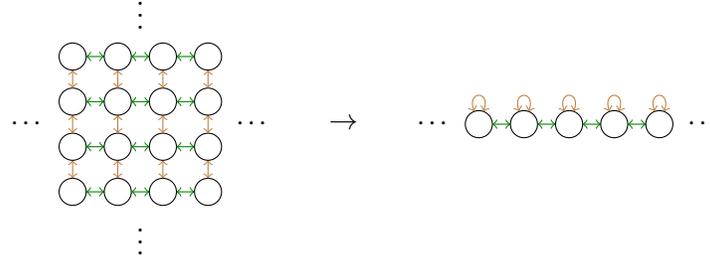

If $k=1$ and $l=0$, we obtain a mapping of the form $\rho(x,y) = x$.
This is a trivial geometrical projection along the $y$ axis.
The corresponding projection operator \eqref{eq:red} acts on a generic $\ket{\psi} \in \hilb$ as
\begin{equation}
  \ket{\psi} = \sum_{x,y\in\mathbb{Z}} \ket{x,y}\otimes\ket{\alpha_{x,y}} : \quad
  \red\ket{\psi} = \sum_{x\in\mathbb{Z}} \left( \ket{x} \otimes \sum_{y\in\mathbb{Z}} \ket{\alpha_{x,y}} \right).
\label{}
\end{equation}
A given quantum walk must have a coin transform homogeneous along the $y$ coordinate to allow this projection, so we consider a coin operator of the form
\begin{equation}
  C : \ket{x,y}\ket{c} \mapsto \ket{x,y} \otimes C_x\ket{c}.
\label{}
\end{equation}
The result of the projection will then be a quantum walk of $\hilb'$ with the coin operator
\begin{equation}
  C' : \ket{x}\ket{c} \mapsto \ket{x} \otimes C_x\ket{c}
\label{}
\end{equation}
and the step operator
\begin{equation}
  \begin{aligned}
    S' :
    &\ket{x}\ket{R} \mapsto \ket{x+1}\ket{R}, \\
    &\ket{x}\ket{L} \mapsto \ket{x-1}\ket{L}, \\
    &\ket{x}\ket{U} \mapsto \ket{x}\ket{U}, \\
    &\ket{x}\ket{D} \mapsto \ket{x}\ket{D},
  \end{aligned}
\label{}
\end{equation}
which is an equivalent form of the operator \eqref{eq:4d-step-2d} used in Sec.~\ref{sec:intro}.
The walking graph of the resulting walk is shown in Fig.~\ref{fig:lazy}.

Note that this step operator admits a step to the right, to the left and no change in the position register, with two coin states corresponding to the latter, and as such represents a lazy quantum walk.
Lazy quantum walks in the literature are often introduced with a three-dimensional coin register~\cite{Inui-lazy}, in which only one basis state corresponds to the no-displacement step.
In order to reduce the above walk to the latter case, one can easily restrict the choice coin operators $C_x$ to those that admit a three-dimensional invariant subspace, for example, $\Span\{\ket{R}, \ket{L}, \ket{U}\}$, and restrict the initial state of the coin register to the same subspace.

\begin{figure}[t]
\centering
\begin{tikzpicture}[scale=0.6,
g1/.style={green!50!black,<->},
g2/.style={brown,<->}]
\path[use as bounding box] (-7.5, -3) rectangle (8.5, 3);
\begin{scope}[shift={(-6,-1.5)}]
\foreach \x in {0,...,3} {
\foreach \y in {0,...,3} {
\draw (\x,\y) circle (0.3);
\ifnum\x<3 \draw[g1] (\x+.3,\y) -- (\x+.7,\y); \fi
\ifnum\y<3 \draw[g2] (\x,\y+.3) -- (\x,\y+.7); \fi
}}
\node at (-.4,1.5) [anchor=east] {$\cdots$};
\node at (3.4,1.5) [anchor=west] {$\cdots$};
\node at (1.5,-.2) [anchor=north] {$\vdots$};
\node at (1.5,3.4) [anchor=south] {$\vdots$};
\foreach \y in {0,...,2}
\draw[red,dashed] (3.3, \y-0.1) -- (-.3, \y+1.1);
\end{scope}
\node at (0,0) {$\rightarrow$};
\begin{scope}[shift={(3,0)}]
\node at (-.4,0) [anchor=east] {$\cdots$};
\node at (4.4,0) [anchor=west] {$\cdots$};
\clip (-.31, -.31) rectangle (4.31,1);
\foreach \x in {0,...,4} {
\draw (\x,0) circle (0.3);
\ifnum\x<4 \draw[g1] (\x+.3,0) -- (\x+.7,0); \fi
}
\foreach \x in {-2,...,3} {
\coordinate (A) at ($(\x,0) + (40:.3)$);
\coordinate (B) at ($(\x+3,0) + (140:.3)$);
\draw[g2] (A) to [bend left=40] (B);
}
\end{scope}
\end{tikzpicture}
\caption{Using a different $\rho$ the 2D quantum walk is mapped to a quantum walk with jumps of constant size.}
\label{fig:jumps}
\end{figure}
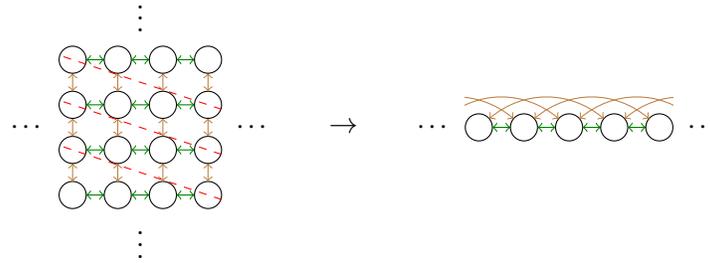

Other interesting choices of $(k,l)$ include $l=1$, $k\in\mathbb{Z}$, which similarly produces a quantum walk on a line with displacements $+1$, $-1$, $+k$, and $-k$.
This is illustrated in Fig.~\ref{fig:jumps}.
With a special case of $k=1$, we obtain a walk on a line with doubled edges.
This is an interesting scenario because it is one the dynamics of which are explored experimentally in \cite{Michelson}.

\begin{figure}[t]
\centering
\begin{tikzpicture}[scale=0.6]
\path[use as bounding box] (-7.5, -3) rectangle (8.5, 3);
\colorlet{c1}{red!50!black}
\colorlet{c2}{green!50!black}
\colorlet{c3}{blue!50!black}
\colorlet{c4}{yellow!50!black}
\begin{scope}[shift={(-3.5,2)}]
\foreach \x in {0,...,6}
\draw[<->] (\x+.3,0) -- (\x+.7,0);
\draw [fill=c1!50!white] (0,0) circle (0.3);
\draw [fill=c2!50!white] (1,0) circle (0.3);
\draw [fill=c3!50!white] (2,0) circle (0.3);
\draw [fill=c4!50!white] (3,0) circle (0.3);
\draw [fill=c1!50!white] (4,0) circle (0.3);
\draw [fill=c2!50!white] (5,0) circle (0.3);
\draw [fill=c3!50!white] (6,0) circle (0.3);
\draw [fill=c4!50!white] (7,0) circle (0.3);
\node at (-.4,0) [anchor=east] {$\cdots$};
\node at (7.4,0) [anchor=west] {$\cdots$};
\end{scope}
\node at (0,1) {$\downarrow$};
\begin{scope}[shift={(0,-1)}]
\node [fill=c1!50!white] (A) at (-1,-1) [draw,circle] {};
\node [fill=c2!50!white] (B) at (1,-1) [draw,circle] {};
\node [fill=c3!50!white] (C) at (1,1) [draw,circle] {};
\node [fill=c4!50!white] (D) at (-1,1) [draw,circle] {};
\draw[<->] (A) to [bend right=20] (B);
\draw[<->] (B) to [bend right=20] (C);
\draw[<->] (C) to [bend right=20] (D);
\draw[<->] (D) to [bend right=20] (A);
\end{scope}
\end{tikzpicture}
\caption{Mapping from a line to a circle graph with $N$ vertices, here $N=4$.}
\label{fig:circle}
\end{figure}
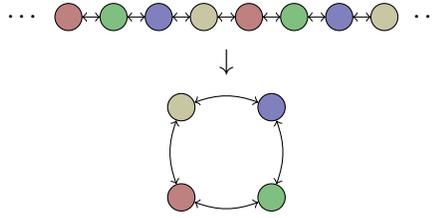

Consider now a quantum walk on a line, where $X = \mathbb{Z}$ and $\Gamma = \{-1, +1\}$.
Coordinate mappings of the above form are not accessible in one dimension, but we can illustrate a ``folding'' operation easily.
Take a fixed $N \in \mathbb{N}$ and the subgroup $H = \langle N \rangle \triangleleft \mathbb{Z}$.
The factor group $\mathbb{Z}/H$ is isomorphic to $\mathbb{Z}_N$, the modular group of remainders of division modulo $N$.
The projection mapping can be formulated as $\rho(x) = x \bmod N$.
The projection operator takes the form
\begin{equation}
  \red: \sum_x \ket{x}\otimes\ket{\alpha_x} \mapsto \sum_{m=0}^{N-1} \left( \ket{m} \otimes \sum_{n\in\mathbb{Z}} \ket{\alpha_{m+nN}} \right),
\label{}
\end{equation}
or in the case of the phase-enhanced operator \eqref{eq:phase}, with the only nontrivial homomorphism (up to a constant multiplication) $\sigma: \mathbb{Z} \to \mathbb{R}: x \mapsto x$,
\begin{equation}
  \red_\varphi: \sum_x \ket{x}\otimes\ket{\alpha_x} \mapsto \sum_{m=0}^{N-1} \left( e^{i\varphi m} \ket{m} \otimes \sum_{n\in\mathbb{Z}} e^{i\varphi nN} \ket{\alpha_{m+nN}} \right).
\label{}
\end{equation}
The result of the projection is a quantum walk on a $N$-circle, optionally with a twisted boundary condition, as illustrated in Fig.~\ref{fig:circle}.
If the latter operator is used, the twist phase is $N\varphi$, it is zero in case of the former.
(Note that any twist phase can be gauged away if the walking graph contains no loops, which is why we did not consider $\red_\varphi$ in the first set of examples.)

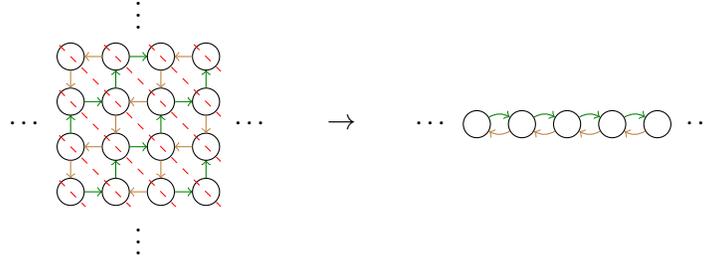
\begin{figure}[t]
\centering
\begin{tikzpicture}[scale=0.6,
g1/.style={green!50!black,->},
g2/.style={brown,->}]
\path[use as bounding box] (-7.5, -3) rectangle (8.5, 3);
\begin{scope}[shift={(-6,-1.5)}]
\foreach \x in {0,...,3} {
\foreach \y in {0,...,3} {
\pgfmathsetmacro\z{\x+\y}
\draw (\x,\y) circle (0.3);
\ifodd\z
\ifnum\y<3 \draw[g1] (\x,\y+.3) -- (\x,\y+.7); \fi
\ifnum\y>0 \draw[g2] (\x,\y-.3) -- (\x,\y-.7); \fi
\else
\ifnum\x<3 \draw[g1] (\x+.3,\y) -- (\x+.7,\y); \fi
\ifnum\x>0 \draw[g2] (\x-.3,\y) -- (\x-.7,\y); \fi
\fi
}}
\node at (-.4,1.5) [anchor=east] {$\cdots$};
\node at (3.4,1.5) [anchor=west] {$\cdots$};
\node at (1.5,-.2) [anchor=north] {$\vdots$};
\node at (1.5,3.4) [anchor=south] {$\vdots$};
\foreach \y in {0,...,2}
\clip (-.31,-.31) rectangle (3.31,3.31);
\foreach \y in {0,...,6}
\draw [red,dashed] (-.5,\y+.5) -- (3.5,\y-3.5);
\end{scope}
\node at (0,0) {$\rightarrow$};
\begin{scope}[shift={(3,0)}]
\node at (-.4,0) [anchor=east] {$\cdots$};
\node at (4.4,0) [anchor=west] {$\cdots$};
\clip (-.31, -.31) rectangle (4.31,1);
\foreach \x in {0,...,4} {
\draw (\x,0) circle (0.3);
\node (A) at (\x,0) [circle] {};
\node (B) at (\x+1,0) [circle] {};
\ifnum\x<4
\draw[g1] (A) to [bend left=30] (B);
\draw[g2] (B) to [bend left=30] (A);
\fi}
\end{scope}
\end{tikzpicture}
\caption{Mapping of the L-lattice to a line graph.}
\label{fig:llat}
\end{figure}

In all the above cases the original walking graph was an Euclidean lattice.
For a more generic (and final) example, consider the following,
\begin{equation}
  G = \langle a,b \mid a^2b^{-2} = e \rangle.
\label{}
\end{equation}
The Cayley graph of this group with the generators $a$ and $b$, and thus also the walking graph if $\Gamma = \{a,b\}$, is the L-lattice, popularized in the quantum walk context by the analogies of the Chalker-Coddington model of the quantum Hall effect \cite{C-C}.
This structure has an interesting projection onto $\mathbb{Z}$ by the homomorphism
\begin{equation}
  \rho: \rho(a) = 1, \rho(b) = -1,
\label{}
\end{equation}
which represents a projection along the diagonals depicted in Fig.~\ref{fig:llat}.
The result is a canonical quantum walk on a line with a two-dimensional coin.
In this case, adding phase information as per \eqref{eq:phase} would be equivalent to merely adjusting local phases on the position state basis of $\hilb_{P'}$ since $\rho$ is also (up to constant multiplication) the only nontrivial homomorphism from $G$ to $(\mathbb{R},+)$.

\section{Applications}

The first possible application of the projection theory is in situations when the quantum walk before projection is better understood than the projected case.
If, in an ultimate scenario, the dynamics is understood analytically in a closed form, or at least asymptotically so, the complete dynamics of the projected walk can be inferred simply using the $\red$ operator.
Many of the examples we have discussed in the previous section start with a simple DTQW on an Euclidean lattice allowing shifts by one along the coordinates.
This is a case that has been studied thoroughly in a great number of works and indeed allows analytical methods, let us mention e.g. \cite{Konno}.

We will illustrate this application on a smaller scale.
It is known that a DTQW on a 2D Euclidean lattice with a homogeneous coin described by the Grover coin matrix,
\begin{equation}
  C = \frac{1}{2} \begin{pmatrix}
  -1 & 1 & 1 & 1 \\
  1 & -1 & 1 & 1 \\
  1 & 1 & -1 & 1 \\
  1 & 1 & 1 & -1
\end{pmatrix},
\label{}
\end{equation}
supports ``trapped states'' which never leave their finite support~\cite{Inui}.
Any part of the initial state that has an overlap with these states will not participate in the otherwise ballistic spreading.
The existence of the trapped states is due to the following pairs of eigenstates of the unitary evolution operator~\cite{Martin}
\begin{equation}
  \begin{aligned}
    \ket{\varphi_{x,y}^\pm} = \frac{1}{2\sqrt{2}} \Big(
      &\ket{x,y}\otimes(\ket{L} + \ket{D})
      \pm \ket{x,y+1}\otimes(\ket{L} + \ket{U}) \\
      &\pm \ket{x+1,y}\otimes(\ket{R} + \ket{D})
      + \ket{x+1,y+1}\otimes(\ket{R} + \ket{U})
    \Big).
  \end{aligned}
\label{}
\end{equation}

From the previous section we know that the 2D DTQW can be projected to, among others, a lazy quantum walk or a quantum walk on a line with double edges.
The projection theorem immediately provides that the trapping effect exists in these scenarios as well, when the same Grover coin matrix is used.
The corresponding trapped states are (prior to optional normalization)
\begin{equation}
  \red\ket{\varphi_{x,y}^\pm} = \frac{1}{2\sqrt{2}} \Big(
    \ket{x}\otimes((1\pm 1)\ket{L} + \ket{D} \pm \ket{U})
    + \ket{x+1}\otimes((1\pm 1)\ket{R} + \ket{U} \pm \ket{D})
  \Big)
\label{}
\end{equation}
for the lazy quantum walk and
\begin{equation}
    \red\ket{\varphi_{x,y}^\pm} = \frac{1}{2\sqrt{2}} \Big(
      \ket{x+y}\otimes(\ket{L} + \ket{D})
      \pm \ket{x+y+1}\otimes(\ket{L} + \ket{R} + \ket{U} + \ket{D})
      + \ket{x+y+2}\otimes(\ket{R} + \ket{U})
    \Big)
\label{}
\end{equation}
for the double line quantum walk.
This saves a lot of effort in comparison with re-deriving these results from the corresponding definitions using techniques similar to~\cite{Martin}.

Finally in this context, we note that projection was already used in one of the author's own earlier works \cite{Nase-hypercube} to reduce the space complexity of a quantum walk-based search algorithm on a hypercube by one qubit, namely projecting a walk on a hypercube of dimension $n+1$ to another on a hypercube of dimension $n$ with self-loops.

\section{Undoing the projection operation}

Utilizing the homomorphism $\rho: X \to X'$, the operators \eqref{eq:red} or $\eqref{eq:phase}$ are manifestly non-invertible.
This is true in all cases except these where $\rho$ itself is injective, which, however, are completely trivial -- both $\red$ and $\red_\varphi$ then only represent a relabelling of the position space basis states from $\ket{x}$ to $\ket{\rho(x)}$ or $e^{i\varphi\sigma(x)} \ket{\rho(x)}$, respectively.
There is little one can do towards undoing the effect of $\red$, however, with the added phase information in $\red_\varphi$, certain projections become invertible.
In this conference paper we only study one simple case and leave a more general discussion to a prepared extended study.

Continuing from the first example of Sec.~\ref{sec:examples}, we take the $X$ and $\Gamma$ corresponding to the two-dimensional Euclidean DTQW and the homomorphism \eqref{eq:rho-k,l}.
As argued therein, we can find $u,v \in \mathbb{Z}$ coprime such that $uk + vl = 1$.
In other words, the matrix
\begin{equation}
M = \begin{pmatrix}
k & l \\ -v & u
\end{pmatrix} \in \mathbb{Z}^{2,2}
\label{}
\end{equation}
has unit determinant and therefore an inverse
\begin{equation}
M^{-1} = \begin{pmatrix}
u & -l \\ v & k
\end{pmatrix} \in \mathbb{Z}^{2,2}.
\label{eq:m-1}
\end{equation}

In addition to $\rho(x,y) = kx + ly$ we take $\sigma(x,y) = uy - vx$.
By the properties of $M$, the pair of values $r = \rho(x,y), s = \sigma(x,y) \in \mathbb{Z}$ uniquely identifies a point $x, y \in \mathbb{Z}$ and can thus be used as a coordinate transform on $X$.
Let an initial state $\ket{\psi_0} \in \mathop{\mathrm{dom}} \red$ be given, then this is also in the domain of $\red_\varphi$ for all $\varphi\in\mathbb{R}$ because the local phase transform in \eqref{eq:phase} does not affect the membership of an sequence in $\ell^p(\mathbb{Z})$ as discussed in Sec.~\ref{sec:theory}.
Thus for any $n \in \mathbb{N}$, the vector function
\begin{equation}
\varphi \mapsto \red_\varphi\ket{\psi_n}
\label{}
\end{equation}
is well-defined for all $\varphi\in\mathbb{R}$ and periodic with a period of $2\pi$.
Taking the decomposition \eqref{eq:ex:alpha}, it can be written as
\begin{equation}
  \varphi \mapsto \sum_{x,y \in \mathbb{Z}} e^{i\varphi\sigma(x,y)} \ket{\rho(x,y)} \otimes \ket{\alpha_{x,y}^{(n)}}
  = \sum_{r,s\in\mathbb{Z}} e^{i\varphi s} \ket{r} \otimes \ket{\alpha_{ur-ls,vr+ks}^{(n)}},
\label{}
\end{equation}
utilizing the coordinate transform $(r,s) = (\rho(x,y), \sigma(x,y))$ and its inverse as given by \eqref{eq:m-1}.

Taking the inverse Fourier series of this function yields
\begin{equation}
  \frac{1}{2\pi}\int_0^{2\pi} e^{-it\varphi} \red_\varphi\ket{\psi_n} d\varphi
  = \sum_{r,s\in\mathbb{Z}} \delta_{s,t} \ket{r} \otimes \ket{\alpha_{ur-ls,vr+ks}^{(n)}}
  = \sum_{r\in\mathbb{Z}} \ket{r} \otimes \ket{\alpha_{ur-lt,vr+kt}^{(n)}}.
\label{}
\end{equation}
Choosing $t = \sigma(x,y)$, the vector $\ket{\alpha_{x,y}^{(n)}}$ can be extracted from this result using the scalar product in the $\hilb_{P'}$ register:
\begin{equation}
  \ket{\alpha_{x,y}^{(n)}} = \bra{\rho(x,y)} \frac{1}{2\pi}\int_0^{2\pi} e^{-i\sigma(x,y)\varphi} \red_\varphi\ket{\psi_n} d\varphi = \frac{1}{2\pi}\int_0^{2\pi} e^{-i\sigma(x,y)\varphi} \bra{\rho(x,y)} \red_\varphi\ket{\psi_n} d\varphi.
\label{}
\end{equation}
Therefore, it is possible to reconstruct the full form of $\ket{\psi_n}$ from its projections under the constant $\rho$ and $\sigma$ as fixed above, if the projected states $\red_\varphi \ket{\psi_n}$ are known for almost every $\varphi\in(0,2\pi)$.

This result alone could play an important role e.g. in experimental scenarios like \cite{Michelson} which are by construction limited to a one-dimensional walking space but have a higher than two-dimensional coin space.
Equation \eqref{eq:s'-phase} shows that the various phases $\varphi$ could be accessed by varying a phase parameter in the step operator, which could perhaps be achieved directly by means of sub-wavelength optical pulse dilations.
However, since the mathematical form of \eqref{eq:s'-phase} can also be written as the operator $S'$ of \eqref{eq:s'} multiplied from the right by a unitary operator diagonal in the whole tensor product basis of $\hilb'$, we can equivalently absorb the local phases into the $C'$ operator.
(Then the commutation relations \eqref{eq:commutation} would be violated but it would remain valid that $\red_\varphi SC = S'C' \red_\varphi$, which is sufficient for the validity of the discussion in this section.)
Such experiment could thus unlock the possibility to observe the full two-dimensional dynamics, if amplitudes rather than probabilities can be measured in it.

\section{Conclusions}

We have shown on a number of examples that an Euclidean quantum walk can, by means of projection, describe the dynamics of various types of quantum walks on lower-dimensional but not necessarily simpler graph structures.
The reach of the projection theory goes beyond these cases, as more complicated graphs can be considered.
Nevertheless, knowing the dynamics, or some peculiar properties, of the original quantum walk, we can learn the same about any of its projections.
We have illustrated this by inferring the presence and mathematical form of trapped states for two interesting types of quantum walks on lines.

The projections of a given quantum walk retain the original coin space and as such often have a higher-dimensional coin than the degree of the projected version of the graph.
This hyper-dimensionality could be taken as a sign that a given quantum walk might itself be a projection of another DTQW on a higher-dimensional structure, and the latter sought for in case methods for solving it are more readily available.
Moreover, the projection mapping in certain cases is reversible, which could mean that the higher-dimensional walk could be completely reconstructed from this projection, given access to freely varying a part of its phase parameters.

The projection theorem can, in summary, be used to simplifying the study of some DTQW problems.
It is not to be confused with other techniques which look superficially similar, namely separating dynamics in two or more orthogonal subspaces 
or using symmetries of the walking graph and the unitary operator to collapse the former to a simpler structure \cite{Childs,SKW}.
These techniques are in fact independent of projection, meaning that the benefits of both could also be combined in solving a given DTQW problem.

\section{Acknowledgements}

This work was supported by the Czech Science Foundation, grant number GA \v{C}R 19-15744Y.

\nocite{*}
\bibliographystyle{eptcs}
\bibliography{potocek}
\end{document}